# Proximal quantum control of spin and spin ensemble with highly localized control field from skyrmions

Md Fahim F Chowdhury[1], Mohamad Niknam[2,4], Md Mahadi Rajib[1], Louis S. Bouchard[2,4], Jayasimha Atulasimha[*,1,3,5,6]

[1]Department of Mechanical and Nuclear Engineering, Virginia Commonwealth University, Richmond, VA, USA.
[2]Department of Chemistry and Biochemistry, University of California Los Angeles, CA, USA.
[3]Department of Electrical and Computer engineering, Virginia Commonwealth University, Richmond, VA, USA.
[4]Center for Quantum Science and Engineering, UCLA, Los Angeles, CA, USA.
[5]Materials Sciences Division. Lawrence Berkeley National Laboratory, Berkeley, CA, USA.
[6]Department of Electrical Engineering and Computer Sciences, University of California, Berkeley, CA, USA.
[*]jatulasimha@vcu.edu

**Abstract**

Selective control of individual spin qubits is needed for scalable quantum computing based on spin states. Achieving high-fidelity in both single and two-qubit gates, essential components of universal quantum computers, necessitates highly localized control fields. These fields must be capable of addressing specific spin qubits while minimizing gate errors and cross-talk in adjacent qubits. Overcoming the challenge of generating a localized radio-frequency magnetic field, in the absence of elementary magnetic monopoles, we introduce a technique that combines divergent and convergent nanoscale magnetic skyrmions. This approach produces a precise control field that manipulates spin qubits with high fidelity. We propose the use of 2D skyrmions, which are 2D analogues of 3D hedgehog structures. The latter are emergent magnetic monopoles, but difficult to fabricate. The 2D skyrmions, on the other hand, can be fabricated using standard semiconductor foundry processes. Our comparative analysis of the density matrix evolution and gate fidelities in scenarios involving proximal skyrmions and nanomagnets indicates potential gate fidelities surpassing 99.95% for $\pi/2$-gates and 99.90% for $\pi$-gates. Notably, the skyrmion configuration generates a significantly lower field on neighboring spin qubits, i.e. 15 times smaller field on a neighboring qubit compared to nanomagnets that produces the same field at the controlled qubit, making it a more suitable candidate for scalable quantum control architectures by reducing disturbances in adjacent qubits.

**Introduction:**

In recent decades, diverse paths in information processing have emerged, exploring various architectures and technologies. These include quantum computing, optical computing, bio-computing using biomolecules such as DNA and proteins, edge computing, and innovations in biological neuron-inspired computing, as well as advancements in traditional von-Neumann computing [1-5]. These developments are often driven by limitations in further miniaturizing CMOS technology and the need for increased switching speeds [6-9]. Additionally, the immense data flow in today's technological arena demands data processing that is not only high in density and speed but also energy-efficient.

Among these, quantum computing could exponentially accelerate computational processes, tackling complex problems beyond the reach of classical computing. Key advancements in this field include the development of algorithms like Shor's and Grover's [9-11]. Quantum computers operate on principles such as superposition, allowing qubits to exist in multiple states simultaneously; entanglement, which interconnects qubits; and quantum interference, where probability amplitudes of quantum states can interfere constructively or destructively. Quantum systems find applications in cryptography, drug discovery, machine learning, communication, sensing, and imaging, among others [12-18]. Various qubit systems have been developed, including superconducting qubits [20-24], spin qubits [25], trapped ions [26], topological qubits [27], and NV centers in diamonds [28], aiming to meet DiVincenzo's criteria for effective qubit implementation. However, challenges such as error-proneness in physical qubits due to relaxation, phase decoherence, thermal noise, and crosstalk during initialization, measurement, and logic gate implementation persist [29-30]. To date, no technology has demonstrated a high number of error-free qubits, a critical requirement for scalability in line with DiVincenzo's criteria.

Single-qubit and two-qubit gates are essential for quantum circuits in universal quantum computers [31-32]. Implementing gates for spin qubits involves generating a control field using microwave or RF sources to rotate qubits. This necessitates RF transmission lines or waveguides for signal delivery [33]. Scaling quantum computers is challenging due to the complexity of integrating and interconnecting a large number of qubits, which requires an even greater number of transmission and readout lines, especially in two-dimensional arrays on substrates. Control pulses in this setup decay spatially, require high current density, and generate joule-heating, potentially leading to qubit decoherence [34]. Spatial inhomogeneities and the non-locality of the field also result in reduced gate fidelity and crosstalk, posing significant barriers to dense qubit architectures [35].

To overcome inhomogeneity, an ideal approach would involve the use of magnetic monopoles for time-varying control field application. Unfortunately, elementary monopoles do not exist. Emergent monopoles such as those emerging through spin ice vertex excitations are of interest, but efficient control remain difficult [36-41]. In the 3D Heisenberg model, spin hedgehogs, topological point defects, display characteristics of emergent magnetic monopoles [42]. They serve as quantized origins and endpoints for gauge fields. However, we are unaware of experimental realizations of 3D hedgehogs, probably because of the inherent difficulty in fabricating them using semiconductor foundry processes. Inspired by their structural similarity to 3D hedgehog configurations, we focus on 2D skyrmions because they can be realized experimentally.

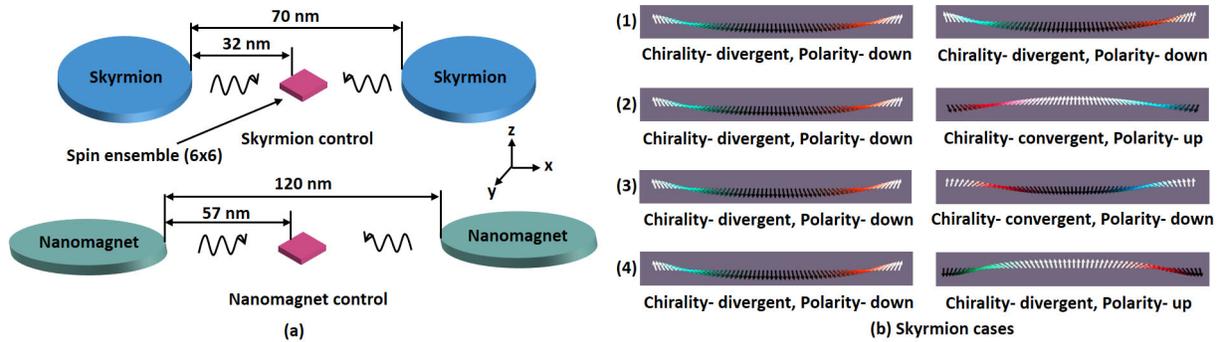

**Figure 1: Schematic diagram of micromagnetic simulation setup in MuMax3. (a) Circular 60 nm skyrmions (top) and elliptical (100 nm x 36 nm) nanomagnets (bottom) to control spin ensemble. (b) Different cases of skyrmions with variable chirality and polarity and the study the magnetic field produced in the spin ensemble area.**

In this study, we evaluate the efficacy of magnetic skyrmions in generating control fields by examining their field homogeneity and the consequent gate fidelity in spin ensembles [43-52]. We envisage a frequency-selective design to achieve spatial selectivity, where qubits are individually addressed at their respective resonant frequency [53-56]. Skyrmions, as topologically protected defects in magnetic thin films, offer good stability at room temperature and can be manipulated with notable energy efficiency. We have analyzed four distinct skyrmion orientations, characterized by what we term as divergent/convergent chirality and up/down polarity, as illustrated in Figure 1(b). From these orientations, configurations 2 and 3 are capable of generating a magnetic field of ample amplitude to manipulate spin ensemble qubits for quantum gate implementation. Spin ensemble qubits have been proposed for roles in quantum entanglement, as bona fide qubits, and in quantum memory and sensing applications [43-51]. Despite the necessity for high gate fidelities, ensemble qubits are subject to lower fidelities due to spatial inhomogeneities. In contrast to nanomagnets, skyrmions create a more localized magnetic field, advantageous for the scalable incorporation of multi-qubit systems. While the average fidelity in the skyrmion scenario is marginally reduced compared to the nanomagnet case, the heightened energy efficiency and localized control with minimal crosstalk in the skyrmion configuration offer a compelling compromise for practical deployment.

**Micromagnetic simulation:**

In a previous study, we demonstrated the feasibility of high-fidelity single-qubit quantum gates using the magnetic field induced by nanomagnets [52]. This field exhibits sufficient homogeneity to facilitate π/2 and π gates in a two-level quantum system, particularly when employing a pair of spintronic devices to target an ensemble qubit centrally located. As depicted in Figure 1, the micromagnetic simulation setup includes both skyrmions and nanomagnets. We have optimized the spacing between the spintronic devices to achieve an induced magnetic field of nearly 1 mT at the center, which is the intended location for the spin ensemble—a 6x6 array comprising 36 spins. We consider four distinct scenarios of skyrmions, differentiated by their chirality and polarity. In each case, we simulated the magnetization dynamics obtained the induced magnetic fields. To simulate magnetization dynamics within the nanomagnets and skyrmions, we solved the Landau-Lifshitz-Gilbert (LLG) equation. These simulations were conducted with a uniform cell size of 1 x 1 x 1 nm³, using the micromagnetic computational framework MuMax3 [57].

$$\frac{d\vec{m}}{dt} = -\frac{\gamma}{(1+\alpha^2)}\left[\left(\vec{m} \times \vec{H}_{eff}\right) + \alpha\left(\vec{m} \times \left(\vec{m} \times \vec{H}_{eff}\right)\right)\right] \quad . \tag{1}$$

Here, $\alpha$ is the Gilbert damping coefficient, $\gamma$ is the gyromagnetic ratio, $\vec{m} = \frac{\vec{M}}{M_s}$ is the normalized magnetization, where $\vec{M}$ is the magnetization and $M_s$ is the saturation magnetization.

The effective magnetic field, $\vec{H}_{eff}$ is sum of exchange field, uniaxial anisotropy field of the nanomagnets, and demagnetizing field:

$$\vec{H}_{eff} = \vec{H}_{anis} + \vec{H}_{exchange} + \vec{H}_{DMI} + \vec{H}_{demag} + \vec{H}_{thermal} \tag{2}$$

$\vec{H}_{anis}$ is the effective field due to uniaxial perpendicular magnetic anisotropy (PMA) which can be modulated using Voltage Control of Magnetic Anisotropy (VCMA) [58-74], $\vec{H}_{exchange}$ is the effective field due to Heisenberg exchange coupling, $\vec{H}_{DMI}$ is the DMI interaction (in the case of a skyrmion), and $\vec{H}_{demag}$ is the field due to the demagnetization energy (shape anisotropy).

In all cases, the skyrmion simulations were initialized to a ferromagnetic state (all spins pointing along the z-direction) and the Néel skyrmions were stabilized at a reduced PMA through VCMA. The PMA was optimized to form a circular skyrmion while keeping other magnetic material parameters constant. The effective PMA energy is expressed as $K_{eff}V = \left(K_{u1} - \frac{1}{2}\mu_0 M_s^2\right)V$, where $K_{eff}$ is the effective PMA energy density, $K_{u1}$ is the PMA constant (uniaxial anisotropy) and V is the volume of the nanodot. We formulate 4 different cases with 2 skyrmions in each case assuming a 6x6 spin ensemble at the centre. The PMA is varied periodically at 500 MHz frequency to manipulate the magnetization of the skyrmions ($m_x, m_y, m_z$) to induce a periodic stray field in the centre to control the spin ensemble. Furthermore, the distance between the skyrmions edges is 70 nm, which induces a ~1 mT field at the spin ensemble.

Unlike the skyrmion case, the nanomagnet needs a bias magnetic field to orient the magnetization in a particular +/-x direction when the shape anisotropy drives the magnetization to the easy (either positive/negative x) axis of the elliptical nanomagnet with equal probability [67]. An exchange bias ($B_{ex}$) can be introduced locally in the ferromagnetic/anti-ferromagnetic interface. The rotation of the magnetization through VCMA induces a magnetic field along the x-axis in the spin ensemble, located at the center of the two nanomagnets.

**Control field induced from the skyrmions and nanomagnets:**

The magnetic field induced in the region of the spin ensemble was calculated when the skyrmions were stabilized in the position shown in Figure 1 (b). The induced $B_x$ was very low in case 1 and 4, which is insufficient to drive the spin ensemble to implement fast high-fidelity gates. In both cases, the divergent skyrmions cancels the effect of the $B_x$-component and produces a very small magnetic field at the center regardless of the polarity of the skyrmions. The amplitude varies from -0.2 mT to 0.2 mT in the ensemble having a positive or negative resultant magnitude in the ensemble qubit according to their relative distance from the skyrmions and reaches ~0 mT at the center. Similar magnetic field characteristics are expected when both skyrmions are convergent. However, when one skyrmion is divergent and the other is convergent, a higher magnetic field amplitude is achieved since both skyrmions have field components adding up along the positive or negative x-axis. The field amplitude was 1 mT for both cases 2 and 3 at the center. The induced field was homogeneous with only 1.2 % deviation in the region of the spin ensemble. Note that if we had considered a single spin qubit, the field gradient would not have been an issue. But we consider the more general case of a spin qubit based on spin ensembles (36 non-interacting spins placed in a 6x6 spin array) to show that even in this more demanding case, quantum control is feasible with minimum disturbance of neighbouring spin qubits.

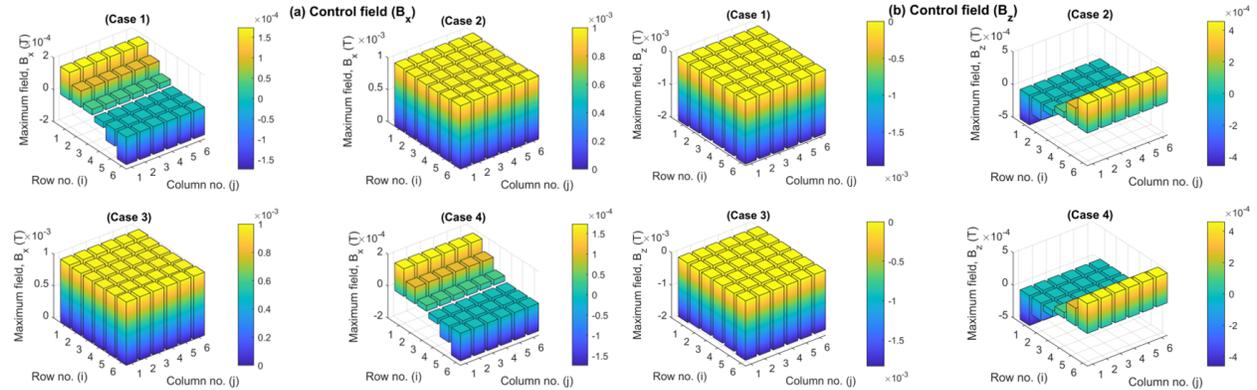

Figure 2: Comparison of (a) x-component ($B_x$) and (b) z-component ($B_z$) of the skyrmion generated magnetic field for 4-cases with different chirality and polarity.

The amplitude and polarity of the induced field along the z-axis, $B_z$, depends on the core-polarities of the skyrmions and affect the rotation frequency of the spin ensemble. In the cases where both skyrmion-cores are downward, there is a resultant magnetic field at the region of the spin ensemble along the negative z-axis with an amplitude of -2 mT. In cases 2 and 4, the skyrmions are in opposite polarity and effectively cancel the effects of stray fields, resulting in a lower induced field, $B_z$. The residual field amplitude along the z-axis varies from -0.05 mT to 0.05 mT depending on the respective distance of the spins to the nearest skyrmions.

The magnetization dynamics of the nanomagnets have been optimized using identical magnetic material parameters as those of the skyrmions to ensure a fair comparison. Both entities, the single nanomagnet and the skyrmion, possess equivalent volumes. The separation between the spintronic devices was set at 70 nm for skyrmions and 110 nm for nanomagnets. In the realm of single-domain nanomagnets with lateral dimensions under 100 nm [75], spin coherence is maintained as they rotate uniformly in response to the application of voltage-controlled magnetic anisotropy (VCMA). In contrast, skyrmions, with their topologically distinct magnetic configuration, exhibit spins that are arranged in a twisted, spiraling, and chiral fashion [76-77]. This unique arrangement results in a reduced stray field from skyrmions when compared to nanomagnets at similar distances. For instance, within a given skyrmion in Figure 1(b), the magnetic moments on opposing sides of the core are antiparallel. This configuration generates an opposing polarity field along the x-axis ($B_x$), which aligns with the direction of the adjacent magnetic moment.

Due to the uniformity of the magnetic moments in nanomagnets, the field they induce is significantly higher than that of the skyrmions, despite both having the same volume. Thus, the inter-device distance for nanomagnets is substantially greater than that for skyrmions to produce an approximate 1 mT field at the spin ensemble location. Detailed simulation parameters for the micromagnetic study are enumerated in Table I.

Table I. Parameter optimization and comparison: Nanomagnets and skyrmion

| Parameter | Skyrmion | Nanomagnet |
| --- | --- | --- |
| Saturation magnetization, $M_s$ | 1.0 x 10$^6$ A/m | 1.0 x 10$^6$ A/m |
| Exchange stiffness, $A_{ex}$ | 15 x 10$^{-12}$ J/m | 15 x 10$^{-12}$ J/m |
| Perpendicular magnetic anisotropy, $ku1$ | 5 x 10$^5$ Jm$^{-3}$ | 1 x 10$^5$ Jm$^{-3}$ |
| Gilbert damping constant, $\alpha$ | 0.01 | 0.01 |
| Thickness, $t$ | 1 nm | 1 nm |
| Easy axis diameter, $D_{easy}$ | 60 nm | 100 nm |
| Hard axis diameter, $D_{hard}$ | 60 nm | 36 nm |
| Volume, $V$ | 900$\pi$ nm$^3$ | 900$\pi$ nm$^3$ |
| Distance | 70 nm | 110 nm |

**Scalability of control field for quantum control of qubits:**

Advancing scalable quantum information processing necessitates densely packed atomic spins within semiconductor devices [78-80]. To scale spin qubit-based quantum computers, precise control over individual spins (or ensemble of spins) is crucial, ensuring that adjacent spins remain unaffected. Control of nuclear and electron spins is traditionally achieved through nuclear magnetic resonance (NMR) and electron spin resonance (EPR). This involves polarizing spins in a static magnetic field and using microwave frequency control pulses corresponding to their Larmor frequencies for spin state manipulation. However, implementing high-fidelity gates requires localized control fields to individually address qubits and prevent crosstalk, which can compromise gate fidelity by inadvertently affecting neighboring qubits [78]. Localizing oscillating control fields to target specific qubits presents challenges, including the wavelength of the RF/microwave field and the complexity of the control circuitry [81].

Recent advancements have shown that electrical control over nuclear spin is possible by employing an electric field from a nanometer-scale electrode tip, allowing access to the spin of single atoms [82-84]. Nuclear electrical resonance (NER) thus introduces the potential for electrically driven qubits. Nevertheless, the high power needed at the antenna may generate thermal noise, leading to decoherence. Magnetic resonance control and detection of spin ensembles is an extensive topic of research across various fields such as chemistry, medicine, biology, material science, and geophysics. Yet, generating a local oscillating magnetic field tailored for quantum computing applications remains a significant hurdle.

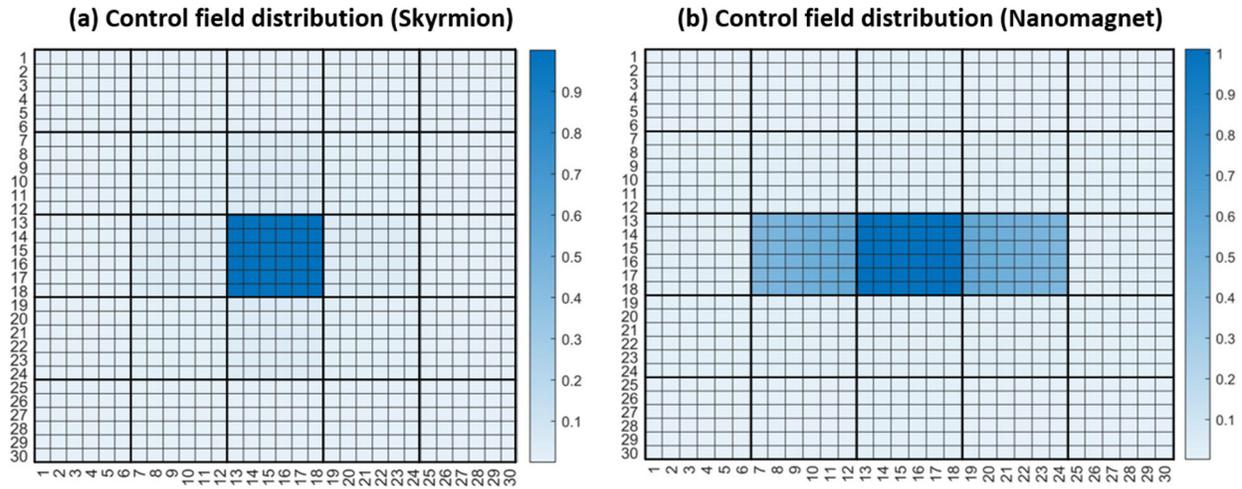

**Figure 3:** Induced magnetic field along the area in the case of (a) skyrmion case2 and (b) nanomagnet. Each square box (bold solid line) represents a 6x6 spin ensemble assumed to be placed periodically 200 nm apart along horizontal-axis and 110 nm along the vertical axis.

To assess the spatial locality of the magnetic fields produced by skyrmions and nanomagnets, we performed numerical calculations of the field amplitude across a two-dimensional plane, assuming the placement of spin ensembles in periodic positions. We considered a 6x6 qubit array with spin ensembles spaced 200 nm apart on the x-axis and 110 nm on the y-axis. The spatial magnetic field's amplitude and the comparison between skyrmion and nanomagnet-induced fields are presented in Figures 3(a,b). For skyrmions, the average normalized amplitude of the induced field at the targeted qubit site is approximately 1.0 mT. The average induced field at the nearest neighboring qubit is 0.0332 mT along the x-axis and 0.0372 mT along the y-axis. In contrast, with nanomagnets, the average amplitude of the magnetic field at the nearest neighbors, particularly in the x-direction, is significantly higher, even though the target spin ensemble experiences a similar magnetic field amplitude of ~1.0 mT. The average induced field at the nearest neighboring qubit on the x-axis is 0.507 mT, which is about 15 times greater than that induced by the skyrmion. Consequently, neighboring spin ensembles must be placed at much greater distances from the target qubit in the nanomagnet scenario to experience an equivalent magnetic field as in the skyrmion case. This suggests that a skyrmion-based system could potentially accommodate a denser array of qubits or spin ensembles within the same area compared to a nanomagnet-based system. Therefore, the skyrmion-generated control field emerges as a more viable option for scalable quantum control architectures, offering reduced crosstalk among qubits.

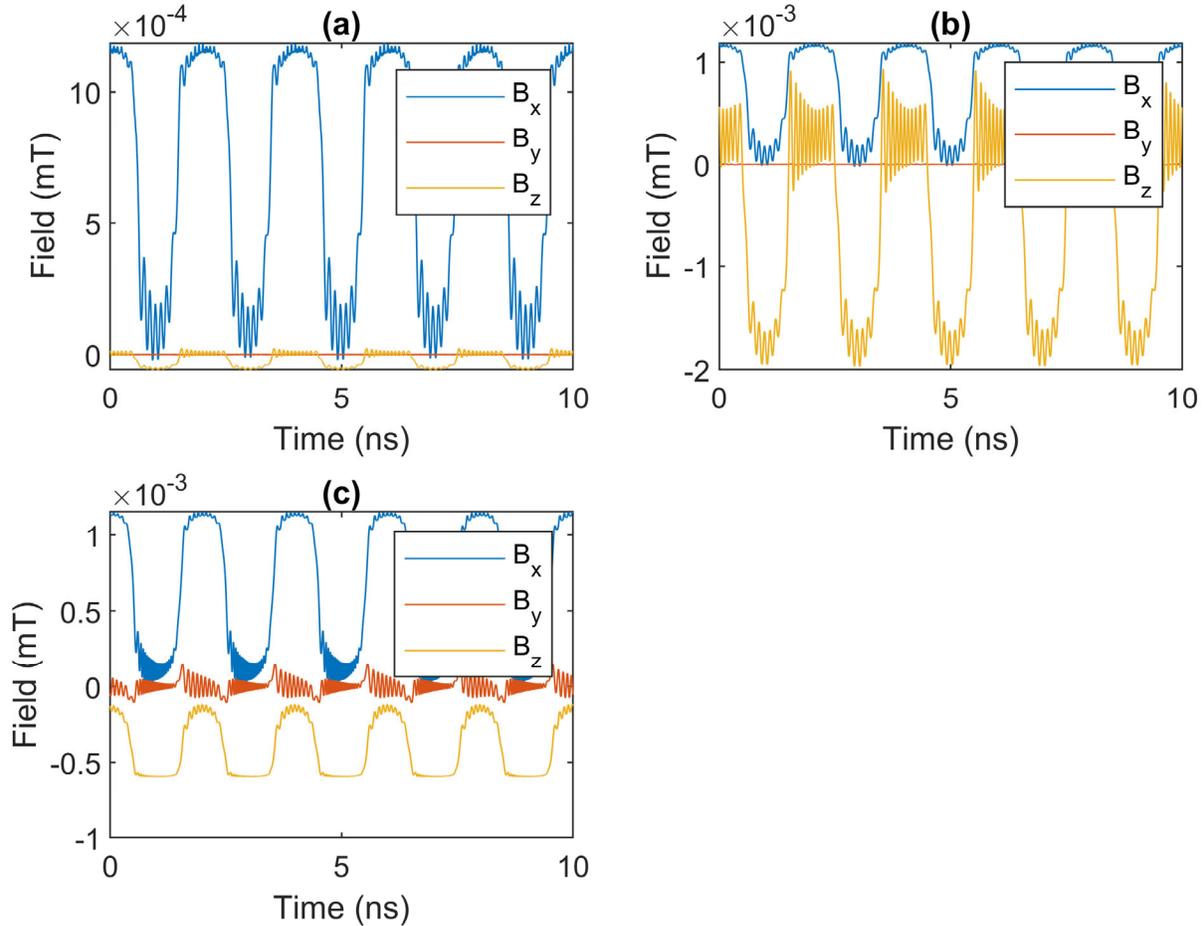

**Figure 4:** Induced magnetic field ($B_x$, $B_y$, $B_z$) by (a) skyrmion case 2 and (b) skyrmion case 3 (c) nanomagnet in spin ensemble in time domain.

    In both scenarios involving skyrmions and nanomagnets, the application of a sinusoidal voltage to induce VCMA results in the generation of a periodic magnetic field. This field, primarily sinusoidal but with higher harmonics due to the nonlinear response of the magnetization to an electric field, is oriented along the x-axis. It interacts with the spins in the mesoscopic qubit [47-56] comprised of the spin ensemble, inducing Larmor precession when the frequency of the field matches the resonance condition. We note that a qubit comprising a single spin would be ideal but read out of a single spin state is considerably harder. Therefore, we consider a mesoscopic qubit consisting of an ensemble of non-interacting (36 spins placed in a 6x6 spin array) for ease of measurement. However, this places a strict constraint on spatial field gradients which can cause dephasing in the ensemble.

The resonance is determined by the effective magnetic bias field in the z-direction, which in turn is linked to the spin's gyromagnetic ratio within the context of a global bias magnetic field. Figure 4 illustrates the microwave fields induced by this process. In the frequency domain, the control fields are characterized by higher harmonic frequencies, with the 500 MHz component being dominant when a 500 MHz VCMA is applied in both skyrmion and nanomagnet cases. These induced fields undergo higher harmonic oscillation during the $90^0$ transitions between in-plane and out-of-plane magnetization. This behavior is attributed to the lower damping constant ($\alpha$), as described by the LLG equation."

**Time evolution of density matrices**

In quantum mechanics, the evolution of a quantum state, $|\psi\rangle$ is governed by the Schrödinger equation,

$$i\frac{\partial |\psi\rangle}{\partial t} = H|\psi\rangle$$

where $H$ is the Hamiltonian operator. The equation of motion of the spin density matrix is the Liouville-Von Neumann equation which is derived from the definition of the statistical operator $\rho(t)$ and the time-dependent Schrodinger equation,

$$i\hbar\frac{\partial}{\partial t}\rho(t) = \sum_i \left(H(t)\rho - \rho H(t)\right) = [H,\rho]$$

$$\frac{\partial \rho(t)}{\partial t} = -\frac{i}{\hbar}[H,\rho]$$

The solution to the Liouville-Von Neumann equation is $\rho(t) = U\rho(t_0)U^\dagger$, which can be described by applying a unitary propagator or time shift operator $U(t,t_0)$,

$$U(t,t_0) = e^{-\frac{i}{\hbar}\int_0^t [H_0 + H_1(t')]dt'}$$

This operator expression allows us to relate the density matrix at a later time $t$ to the density matrix at some earlier time $t_0$,

$$\rho(t) = U(t,t_0)\rho(t_0)U^\dagger(t,t_0)$$

For simplicity, we study the spin dynamics of a two-level quantum system. Spin dynamics in magnetic resonance experiments are governed by the time-dependent component of the Hamiltonian which consists of the induced (internal) and generated (external) microwave fields and applied static field in a closed quantum system. In this case, the Hamiltonian of the spin ensemble is determined by the static field ($B_0$) and induced field of nanomagnets or skyrmions.

The Hamiltonian is:

$$H(t) = -\vec{\mu}.\vec{B_T}(t) = -\gamma\vec{S}.\vec{B_T}(t) = -\gamma_e\vec{S}.\left(B_0 + \vec{B}_{ctrl}(t)\right)$$

where $\vec{\mu}$ is the magnetic moment of the spin and $\vec{B_T}(t)$ is the total magnetic field in the closed system which consists of the static field, $B_0$ and perturbing field, $\vec{B}_{ctrl}(t)$. The magnetic moment, $\vec{\mu} = \gamma_e\vec{S}$ is proportional to its spin angular momentum $\vec{S}$. The gyromagnetic ratio, $\gamma$ is calculated as: $\gamma_e = \frac{eg_e}{2m_e}$, where electron g-factor, $g_e$ =2.00232.

For the depiction of spin dynamics, the electron spin is initialized along the x, y, and z-axes and projected along the x, z, and y-axes after its rotation, respectively. The unitary operator is calculated each time-step followed by Hamiltonian consisting the continuous application of low amplitude RF source induced from the nanomagnets and skyrmions. The spins in the ensemble qubit precess around the effective magnetic field with Larmor frequency which is described by the density matrix ($\rho$). Figure 5 shows the average observed signal ($\rho_x, \rho_y, \rho_z$) for skyrmion and nanomagnet cases at 500 MHz frequency. We observe x-rotations related to the time-dependent component of $\vec{B}_{ctrl}$ along x-axis. To excite the electron spin to produce Rabi oscillation, the drive frequency has to be equal the Larmor frequency. In Figure 5(a), the skyrmion case 2 has a dip indicating an incomplete Rabi oscillation which can lead to lower gate fidelities, but not in the nanomagnet case and skyrmion case 3. This is due to the highly inhomogeneous and opposite polarity $B_z$ field (Figure 2) in ensemble qubit, which affects the rotating field correction at each spin.

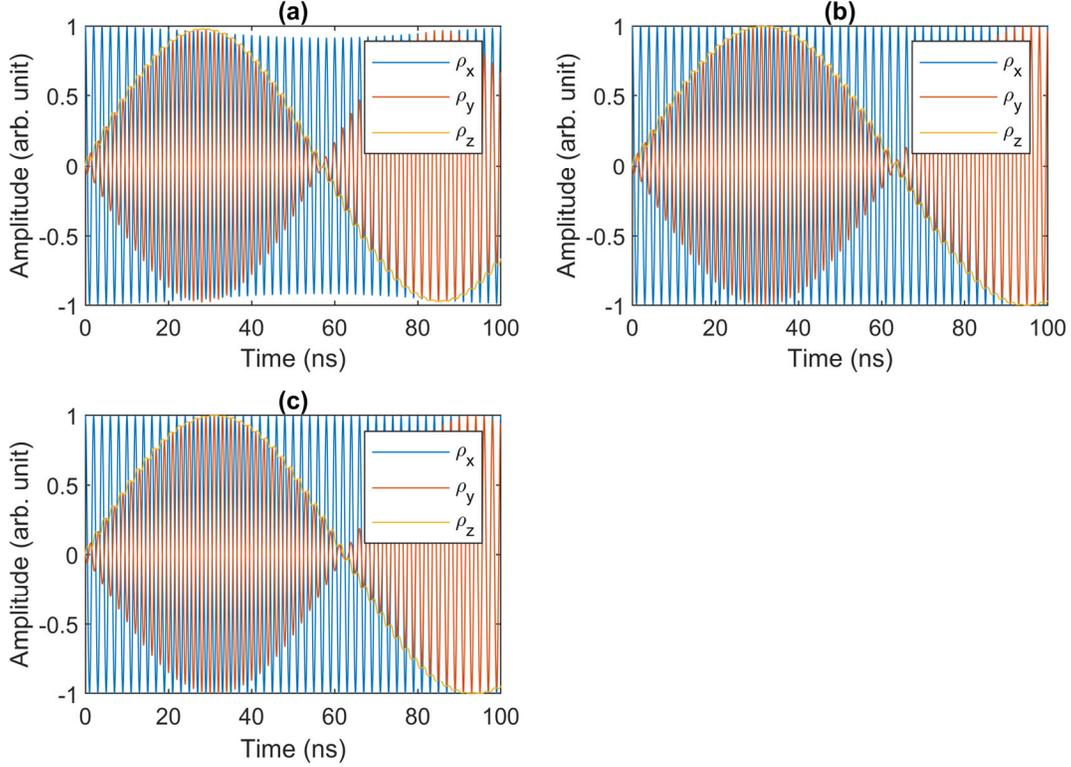

**Figure 5:** Evolution of 36 spins in the lab frame with the induced field at 500 MHz drive with (a) skyrmion case2, (b) skyrmion case3, and (c) nanomagnet.

**Gate fidelities**

Current NISQ quantum systems are prone to decoherence, environmental interference, crosstalk including the measurement errors, noise control circuitry, fabrication defects which lead to lower gate fidelities. Gate fidelity is the capacity of the qubits to maintain their quantum states or execution of correct logics, which also indicates the reliability in computation and critical for algorithm design and applications. The gate fidelity is calculated from:

$$F = \frac{1}{2} + \frac{1}{3} \Sigma_{j=x,y,z} \, Tr[U \frac{\sigma_j}{2} U^j M\left(\frac{\sigma_j}{2}\right)].$$

As discussed earlier, if we had considered a single spin qubit, the field gradient would not have been an issue. However, we consider the more demanding case of a mesoscopic qubit consisting of a spin ensemble with a finite volume. Hence, the fidelity will degrade due to the spatial variation of the control field. The effect of the inhomogeneities is estimated by calculating the fidelities over space as a function of the position followed by averaging over the spin ensemble. Note that for a qubit consisting of a single spin, the fidelity will be better than for the ensemble qubit.

The X/2 gate refers to a rotation of a qubit around x-axis by an angle equal to half of the Bloch sphere's rotation for an X gate (Pauli-X gate). This is also known as the $\pi/2$ pulse or Hadamard gate when applied to a computational basis state and can be achieved by stopping the drive when $\rho_y$ rotates to z, or $\rho_z$ rotates to y. The Pauli-X gate refers to a $\pi$ rotation around the x-axis of the Bloch sphere, which flips the state $|0\rangle$ and $|1\rangle$, while leaving the superposition states on the equator unaffected.

The X/2 and X-gate durations and fidelities are given in Table II. Although, the gate duration is smaller, the gate fidelities for skyrmion case 2 are much lower compared to the nanomagnet case due its significant spatial inhomogeneities in $B_z$ field. But skyrmion case 3 features much higher gate fidelities than both the nanomagnet and skyrmion case 2, and comparable gate duration for both X/2 and X gates. While single spin fidelities ($\pi/2$ avg – 99.99 and $\pi$ avg- 99.95) are much higher than the ensemble average ($\pi/2$ avg – 99.99 and $\pi$ avg- 99.82), the ensemble gives the benefit of measurement and further error correction scheme. To be compatible with the NISQ era, the gate fidelities can be further improved by applying composite pulses such as the Knill pulse, or dynamic decoupling schemes.

Table II: Fidelity with skyrmions and nanomagnet generated control field

|  | Nanomagnet | Skyrmion case 2 | Skyrmion case 3 |
| --- | --- | --- | --- |
| $\pi/2$ max (%) | 99.961 | 99.918 | 99.999 |
| $\pi/2$ avg (%) | 99.948 | 99.286 | 99.985 |
| $T_{\pi/2}$ (ns) | 32.016 | 30.044 | 32.034 |
| $\pi$ max (%) | 99.904 | 99.965 | 99.945 |
| $\pi$ avg (%) | 99.889 | 97.158 | 99.820 |
| $T_\pi$ (ns) | 62.008 | 58.026 | 64.026 |

One of the reasons spin qubits are promising candidates for large-scale quantum computation is because electron spins in silicon can be operated at temperatures well above the ~mK range. Recently, single-qubit control above 1 K with silicon quantum dots and two-qubit logic in a quantum circuit was demonstrated at 1.1 K, which shows that quantum coherence phenomena of spin qubits in the range of T=0.45-1.25 K are well suited for the implementation of spin qubit gates [86-87]. To study the effect of thermal fluctuations on the magnetization dynamics of the nanomagnet controlling such spin qubits, we performed micromagnetic simulations at 1K and compared the results to magnetization dynamics at 0 K. The qualitative deviation of the magnetization and hence the control magnetic field due to the thermal fluctuation at 1K was found to be negligible.

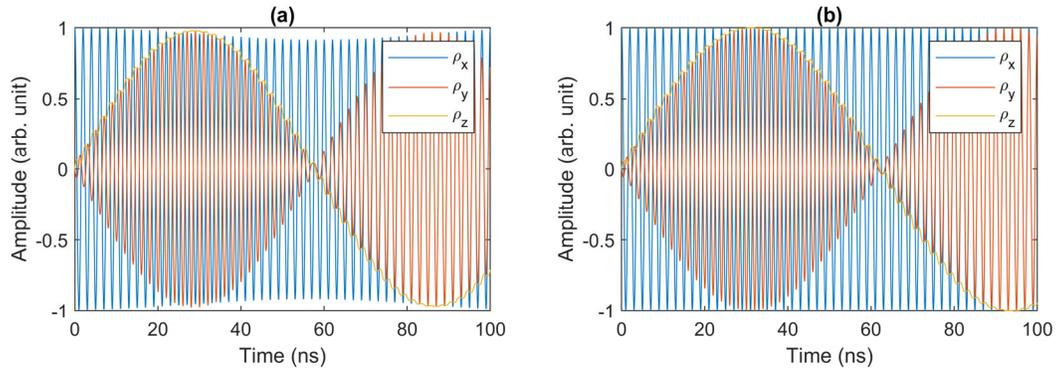

**Figure 6:** Evolution of spins in the lab frame with the induced field at 500 MHz drive with (a) skyrmion case2, (b) nanomagnet subjected to thermal noise.

Furthermore, the $\frac{\pi}{2}$ and $\pi$ gate fidelities for the nanomagnet case are not impacted by the thermal noise at 1 K. While in the skyrmion case the average $\frac{\pi}{2}$-gate fidelity is the same, the average $\pi$-gate fidelity is slightly lower (97.11) than the 0 K fidelity (97.15).

**Conclusion**

The combination of divergent and convergent magnetic skyrmions generates a high magnetic field at the center of the skyrmion pair that is independent of their polarities. The unique topological arrangement of spins within a skyrmion results in a reduced stray field at nearby points compared to that induced by nanomagnets. Consequently, the magnetic field produced by skyrmions is much more localized, a beneficial attribute for RF control fields in systems with multiple qubits separated by short distances. Additionally, our results indicate that maximum gate fidelities surpass 99.95% for π/2-gates and 99.90% for π-gates for both skyrmion and nanomagnet configurations controlling qubits comprising spin ensembles. Fidelities will be better for single spin qubits In conclusion, skyrmions have demonstrated the capability to generate a homogeneous control field that can precisely address spin ensembles with high fidelity. The localized control field achieved in this study is precise enough to manipulate individual qubits without interfering with neighboring ones, a critical requirement for the controlled-X gates that are essential for constructing a scalable quantum computer.

**References:**

1. Xu, X. Y., & Jin, X. M. (2023). Integrated Photonic Computing beyond the von Neumann Architecture. *ACS Photonics*, *10*(4), 1027-1036.
2. Sebastian, A., Le Gallo, M., & Eleftheriou, E. (2019). Computational phase-change memory: Beyond von Neumann computing. *Journal of Physics D: Applied Physics*, *52*(44), 443002.
3. Aspray, W. (1990). *John von Neumann and the origins of modern computing*. Mit Press.
4. Wright, C. D., Hosseini, P., & Diosdado, J. A. V. (2013). Beyond von-Neumann computing with nanoscale phase-change memory devices. *Advanced Functional Materials*, *23*(18), 2248-2254.
5. Ge, R., Wu, X., Kim, M., Shi, J., Sonde, S., Tao, L., ... & Akinwande, D. (2018). Atomristor: nonvolatile resistance switching in atomic sheets of transition metal dichalcogenides. *Nano letters*, *18*(1), 434-441.
6. Sangwan, V. K., & Hersam, M. C. (2020). Neuromorphic nanoelectronic materials. *Nature nanotechnology*, *15*(7), 517-528.
7. Sebastian, A., Le Gallo, M., Khaddam-Aljameh, R., & Eleftheriou, E. (2020). Memory devices and applications for in-memory computing. *Nature nanotechnology*, *15*(7), 529-544.
8. Liu, C., Chen, H., Wang, S., Liu, Q., Jiang, Y. G., Zhang, D. W., ... & Zhou, P. (2020). Two-dimensional materials for next-generation computing technologies. *Nature Nanotechnology*, *15*(7), 545-557.

9. Montanaro, A. (2016). Quantum algorithms: an overview. *npj Quantum Information*, *2*(1), 1-8.
10. Cleve, R., Ekert, A., Macchiavello, C., & Mosca, M. (1998). Quantum algorithms revisited. *Proceedings of the Royal Society of London. Series A: Mathematical, Physical and Engineering Sciences*, *454*(1969), 339-354.
11. Bharti, K., Cervera-Lierta, A., Kyaw, T. H., Haug, T., Alperin-Lea, S., Anand, A., ... & Aspuru-Guzik, A. (2022). Noisy intermediate-scale quantum algorithms. *Reviews of Modern Physics*, *94*(1), 015004.
12. Bayerstadler, A., Becquin, G., Binder, J., Botter, T., Ehm, H., Ehmer, T., Erdmann, M., Gaus, N., Harbach, P., Hess, M. and Klepsch, J., 2021. Industry quantum computing applications. *EPJ Quantum Technology*, *8*(1), p.25.
13. Orús, Román, Samuel Mugel, and Enrique Lizaso. "Quantum computing for finance: Overview and prospects." *Reviews in Physics* 4 (2019): 100028.
14. Preskill, John. "Quantum computing in the NISQ era and beyond." *Quantum* 2 (2018): 79.
15. Budde, Florian, and Daniel Volz. "The next big thing? Quantum computing's potential impact on chemicals." *McKinsey article* (2019).
16. Freeman AJ. Materials by design and the exciting role of quantum computation/simulation. J Comput Appl Math. 2002;149(1):27–56.
17. Gisin, Nicolas, Grégoire Ribordy, Wolfgang Tittel, and Hugo Zbinden. "Quantum cryptography." *Reviews of modern physics* 74, no. 1 (2002): 145.
18. Orús, Román, Samuel Mugel, and Enrique Lizaso. "Quantum computing for finance: Overview and prospects." *Reviews in Physics* 4 (2019): 100028.
19. Steffen, M., DiVincenzo, D. P., Chow, J. M., Theis, T. N., & Ketchen, M. B. (2011). Quantum computing: An IBM perspective. *IBM Journal of Research and Development*, *55*(5), 13-1.
20. Devoret, M. H., Wallraff, A., & Martinis, J. M. (2004). Superconducting qubits: A short review. *arXiv preprint cond-mat/0411174*.
21. Kjaergaard, M., Schwartz, M. E., Braumüller, J., Krantz, P., Wang, J. I. J., Gustavsson, S., & Oliver, W. D. (2020). Superconducting qubits: Current state of play. *Annual Review of Condensed Matter Physics*, *11*, 369-395.
22. Arute, F., Arya, K., Babbush, R., Bacon, D., Bardin, J. C., Barends, R., ... & Martinis, J. M. (2019). Quantum supremacy using a programmable superconducting processor. *Nature*, *574*(7779), 505-510.
23. Krantz, P., Kjaergaard, M., Yan, F., Orlando, T. P., Gustavsson, S., & Oliver, W. D. (2019). A quantum engineer's guide to superconducting qubits. *Applied physics reviews*, *6*(2).
24. Kjaergaard, M., Schwartz, M. E., Braumüller, J., Krantz, P., Wang, J. I. J., Gustavsson, S., & Oliver, W. D. (2020). Superconducting qubits: Current state of play. *Annual Review of Condensed Matter Physics*, *11*, 369-395.
25. Madzik, M. T. et al. Precision tomography of a three-qubit donor quantum processor in silicon. *Nature* **601**, 348–353 (2022).
26. Monroe, C. et al. Programmable quantum simulations of spin systems with trapped ions. *Rev. Mod. Phys.* **93**, 025001 (2021).
27. Lutchyn, R. M. et al. Majorana zero modes in superconductor–semiconductor heterostructures. *Nat. Rev. Mater.* **3**, 52–68 (2018).
28. Pezzagna, S. & Meijer, J. Quantum computer based on color centers in diamond. *Appl Physi. Rev.* **8**, 011308 (2021).
29. Aharonov, D., Kitaev, A., & Nisan, N. (1998, May). Quantum circuits with mixed states. In *Proceedings of the thirtieth annual ACM symposium on Theory of computing* (pp. 20-30).
30. Gu, X., Kockum, A. F., Miranowicz, A., Liu, Y. X., & Nori, F. (2017). Microwave photonics with superconducting quantum circuits. *Physics Reports*, *718*, 1-102.
31. Sete, E. A., Zeng, W. J., & Rigetti, C. T. (2016, October). A functional architecture for scalable quantum computing. In *2016 IEEE International Conference on Rebooting Computing (ICRC)* (pp. 1-6). IEEE.
32. Linke, N. M., Maslov, D., Roetteler, M., Debnath, S., Figgatt, C., Landsman, K. A., ... & Monroe, C. (2017). Experimental comparison of two quantum computing architectures. *Proceedings of the National Academy of Sciences*, *114*(13), 3305-3310.

33. Bardin, J. C., Slichter, D. H., & Reilly, D. J. (2021). Microwaves in quantum computing. *IEEE journal of microwaves*, *1*(1), 403-427.
34. Wang, X., Xiao, Y., Liu, C., Lee-Wong, E., McLaughlin, N. J., Wang, H., ... & Du, C. R. (2020). Electrical control of coherent spin rotation of a single-spin qubit. *npj Quantum Information*, *6*(1), 78.
35. Córcoles, A. D., Kandala, A., Javadi-Abhari, A., McClure, D. T., Cross, A. W., Temme, K., ... & Gambetta, J. M. (2019). Challenges and opportunities of near-term quantum computing systems. *Proceedings of the IEEE*, *108*(8), 1338-1352.
36. Diamantini, M.C., Trugenberger, C.A. and Vinokur, V.M., 2021. Quantum magnetic monopole condensate. *Communications Physics*, *4*(1), p.25.
37. Castelnovo, C., Moessner, R., & Sondhi, S. L. (2008). Magnetic monopoles in spin ice. *Nature*, *451*(7174), 42-45.
38. Anand, N., Barry, K., Neu, J.N., Graf, D.E., Huang, Q., Zhou, H., Siegrist, T., Changlani, H.J. and Beekman, C., 2022. Investigation of the monopole magneto-chemical potential in spin ices using capacitive torque magnetometry. *Nature Communications*, *13*(1), p.3818.
39. Khomskii, D. I. "Electric activity at magnetic moment fragmentation in spin ice." *Nature Communications* 12, no. 1 (2021): 3047.
40. Keswani, N., Lopes, R.J., Nakajima, Y., Singh, R., Chauhan, N., Som, T., Kumar, D.S., Pereira, A.R. and Das, P., 2021. Controlled creation and annihilation of isolated robust emergent magnetic monopole like charged vertices in square artificial spin ice. *Scientific Reports*, *11*(1), p.13593.
41. Paulsen, C., S. R. Giblin, Elsa Lhotel, D. Prabhakaran, K. Matsuhira, G. Balakrishnan, and S. T. Bramwell. "Nuclear spin assisted quantum tunnelling of magnetic monopoles in spin ice." *Nature communications* 10, no. 1 (2019): 1509.
42. Kanazawa, N., Kitaori, A., White, J. S., Ukleev, V., Rønnow, H. M., Tsukazaki, A., ... & Tokura, Y. (2020). Direct observation of the statics and dynamics of emergent magnetic monopoles in a chiral magnet. *Physical review letters*, *125*(13), 137202.
43. Mirkamali, M. S. & Cory, D. G. Mesoscopic spin systems as quantum entanglers. *Phys. Rev. A.* **101**, 032320 (2020).
44. Zhukov, A. A., Shapiro, D. S., Pogosov, W. V. & Lozovik, Y. E. Dynamics of a mesoscopic qubit ensemble coupled to a cavity: Role of collective dark states. *Phys. Rev. A.* **96**, 033804 (2017).
45. Barbara, B. Mesoscopic systems: classical irreversibility and quantum coherence. *Philosophical Transactions of the Royal Society A: Mathematical, Physical and Engineering Sci.* **370**, 4487–4516 (2012).
46. Gangloff, D. A. et al. Witnessing quantum correlations in a nuclear ensemble via an electron spin qubit. *Nat. Phys.* **17**, 1247–1253 (2021).
47. Giedke, G., Taylor, J. M., D'Alessandro, D., Lukin, M. D. & Imamoğlu, A. Quantum measurement of a mesoscopic spin ensemble. *Phys. Rev. A.* **74**, 032316 (2006).
48. Beterov, I. I. et al. Coherent control of mesoscopic atomic ensembles for quantum information. *Laser Phys.* **24**, 074013 (2014).
49. Gangloff, D. A. et al. Witnessing quantum correlations in a nuclear spin ensemble via a proxy qubit. In Quantum Information and Measurement VI 2021, Tu3A.2 (Optica Publishing Group, 2021).
50. Jackson, D. M. et al. Quantum sensing of a coherent single spin excitation in a nuclear ensemble. *Nature Physics* **17**, 585–590 (2021).
51. Rabl, P. et al. Hybrid quantum processors: Molecular ensembles as quantum memory for solid state circuits. *Phys. Rev. Lett.* **97**, 033003 (2006).
52. Niknam, M., Chowdhury, M.F.F., Rajib, M.M., Misba, W.A., Schwartz, R.N., Wang, K.L., Atulasimha, J. and Bouchard, L.S., 2022. Quantum control of spin qubits using nanomagnets. Communications Physics, 5(1), p.284.
53. Hirohata, A., & Takanashi, K. (2014). Future perspectives for spintronic devices. Journal of Physics D: Applied Physics, 47(19), 193001.
54. Dieny, B., Prejbeanu, I. L., Garello, K., Gambardella, P., Freitas, P., Lehndorff, R., ... & Bortolotti, P. (2020). Opportunities and challenges for spintronics in the microelectronics industry. Nature Electronics, 3(8), 446-459.

55. Hirohata, A., Yamada, K., Nakatani, Y., Prejbeanu, I. L., Diény, B., Pirro, P., & Hillebrands, B. (2020). Review on spintronics: Principles and device applications. Journal of Magnetism and Magnetic Materials, 509, 166711.
56. Barla, P., Joshi, V. K., & Bhat, S. (2021). Spintronic devices: a promising alternative to CMOS devices. Journal of Computational Electronics, 20(2), 805-837.
57. Vansteenkiste, A., Leliaert, J., Dvornik, M., Helsen, M., Garcia-Sanchez, F., & Van Waeyenberge, B. (2014). The design and verification of MuMax3. AIP advances, 4(10).
58. Yang, H. X. et al. First-principles investigation of the very large perpendicular magnetic anisotropy at fe|mgo and co|mgo interfaces. Phys. Rev. B. 84, 054401 (2011).
59. Niranjan, M. K., Duan, C.-G., Jaswal, S. S. & Tsymbal, E. Y. Electric field effect on magnetization at the fe/mgo(001) interface. App. Phys.Lett. 96, 222504 (2010).
60. AMIRI, P. K. & WANG, K. L. Voltage-controlled magnetic anisotropy in spintronic devices. SPIN 02, 1240002 (2012).
61. Wang, W.-G., Li, M., Hageman, S. & Chien, C. Electric-field-assisted switching in magnetic tunnel junctions. Nat. Mater. 11, 64–68 (2012).
62. Li, X., Lee, A., Razavi, S. A., Wu, H. & Wang, K. L. Voltage-controlled magnetoelectric memory and logic devices. MRS Bulletin. 43, 970–977 (2018).
63. Bhattacharya, D., Razavi, S. A., Wu, H., Dai, B., Wang, K. L., & Atulasimha, J. (2020). Creation and annihilation of non-volatile fixed magnetic skyrmions using voltage control of magnetic anisotropy. Nature Electronics, 3(9), 539-545.
64. Wang, K. L., Lee, H., & Amiri, P. K. (2015). Magnetoelectric random access memory-based circuit design by using voltage-controlled magnetic anisotropy in magnetic tunnel junctions. IEEE Transactions on Nanotechnology, 14(6), 992-997.
65. Bhattacharya, D., Al-Rashid, M. M., & Atulasimha, J. (2016). Voltage controlled core reversal of fixed magnetic skyrmions without a magnetic field. Scientific reports, 6(1), 31272.
66. Biswas, A. K., Bandyopadhyay, S., & Atulasimha, J. (2014). Complete magnetization reversal in a magnetostrictive nanomagnet with voltage-generated stress: A reliable energy-efficient non-volatile magneto-elastic memory. Applied Physics Letters, 105(7).
67. Bhattacharya, D. & Atulasimha, J. Skyrmion-mediated voltage-controlled switching of ferromagnets for reliable and energy-efficient two-terminal memory. ACS Applied Materials & Inter. 10, 17455–17462 (2018).
68. Rajib, M. M., Misba, W. A., Bhattacharya, D., & Atulasimha, J. (2021). Robust skyrmion mediated reversal of ferromagnetic nanodots of 20 nm lateral dimension with high Ms and observable DMI. *Scientific reports*, *11*(1), 20914.
69. Rajib, M. M., Al Misba, W., Bhattacharya, D., Garcia-Sanchez, F., & Atulasimha, J. (2020). Dynamic skyrmion-mediated switching of perpendicular mtjs: Feasibility analysis of scaling to 20 nm with thermal noise. *IEEE Transactions on Electron Devices*, *67*(9), 3883-3888.
70. Biswas, A. K., Ahmad, H., Atulasimha, J., & Bandyopadhyay, S. (2017). Experimental demonstration of complete 180 reversal of magnetization in isolated Co nanomagnets on a PMN–PT substrate with voltage generated strain. *Nano letters*, *17*(6), 3478-3484.
71. D'Souza, N., Salehi Fashami, M., Bandyopadhyay, S., & Atulasimha, J. (2016). Experimental clocking of nanomagnets with strain for ultralow power Boolean logic. *Nano letters*, *16*(2), 1069-1075.
72. Sampath, V., D'Souza, N., Bhattacharya, D., Atkinson, G. M., Bandyopadhyay, S., & Atulasimha, J. (2016). Acoustic-wave-induced magnetization switching of magnetostrictive nanomagnets from single-domain to nonvolatile vortex states. *Nano Letters*, *16*(9), 5681-5687.
73. Atulasimha, J., & Bandyopadhyay, S. (2010). Bennett clocking of nanomagnetic logic using multiferroic single-domain nanomagnets. *Applied Physics Letters*, *97*(17).
74. Buzzi, M., Chopdekar, R. V., Hockel, J. L., Bur, A., Wu, T., Pilet, N., ... & Nolting, F. (2013). Single domain spin manipulation by electric fields in strain coupled artificial multiferroic nanostructures. *Physical review letters*, *111*(2), 027204.
75. Cowburn, R. P., Koltsov, D. K., Adeyeye, A. O., Welland, M. E., & Tricker, D. M. (1999). Single-domain circular nanomagnets. Physical Review Letters, 83(5), 1042.

76. Finocchio, G., Büttner, F., Tomasello, R., Carpentieri, M., & Kläui, M. (2016). Magnetic skyrmions: from fundamental to applications. Journal of Physics D: Applied Physics, 49(42), 423001.
77. Tokura, Y., & Kanazawa, N. (2020). Magnetic skyrmion materials. Chemical Reviews, 121(5), 2857-2897.
78. Jelezko, F. et al. Observation of coherent oscillation of a single nuclear spin and realization of a two-qubit conditional quantum gate. Phys. Rev. Lett. 93, 130501 (2004).
79. Pla, J. J. et al. High-fidelity readout and control of a nuclear spin qubit in silicon. Nature 496, 334–338 (2013).
80. Willke, P. et al. Hyperfine interaction of individual atoms on a surface. Science 362, 336–339 (2018).
81. Srinivas, R., Löschnauer, C. M., Malinowski, M., Hughes, A. C., Nourshargh, R., Negnevitsky, V., ... & Ballance, C. J. (2023). Coherent Control of Trapped-Ion Qubits with Localized Electric Fields. Physical Review Letters, 131(2), 020601.
82. Asaad, Serwan, Vincent Mourik, Benjamin Joecker, Mark AI Johnson, Andrew D. Baczewski, Hannes R. Firgau, Mateusz T. Mądzik et al. "Coherent electrical control of a single high-spin nucleus in silicon." Nature 579, no. 7798 (2020): 205-209.
83. Yoneda, J. et al. A quantum-dot spin qubit with coherence limited by charge noise and fidelity higher than 99.9%. Nat. Nanotechnol. 13, 102–106 (2018).
84. Dixon, R. W., and N. Bloembergen. "Electrically induced perturbations of halogen nuclear quadrupole interactions in polycrystalline compounds. II. microscopic theory." The Journal of Chemical Physics 41, no. 6 (1964): 1739-1747.
85. Ono, Masaaki, Jun Ishihara, Genki Sato, Yuzo Ohno, and Hideo Ohno. "Coherent manipulation of nuclear spins in semiconductors with an electric field." Applied Physics Express 6, no. 3 (2013): 033002.
86. Yang, C.H., Leon, R.C.C., Hwang, J.C.C., Saraiva, A., Tanttu, T., Huang, W., Camirand Lemyre, J., Chan, K.W., Tan, K.Y., Hudson, F.E. and Itoh, K.M., 2020. Operation of a silicon quantum processor unit cell above one kelvin. Nature, 580(7803), pp.350-354.
87. Petit, L., Eenink, H.G.J., Russ, M., Lawrie, W.I.L., Hendrickx, N.W., Philips, S.G.J., Clarke, J.S., Vandersypen, L.M.K. and Veldhorst, M., 2020. Universal quantum logic in hot silicon qubits. Nature, 580(7803), pp.355-359.

# ACKNOWLEDGEMENT:

All authors were supported by the US National Science Foundation (NSF) ExpandQISE grant # 2231356 and M.M.R and J.A were also supported by NSF SHF: Small grant # 1909030.

# Supplementary

**Micromagnetic simulation:**

The effective field due to the Heisenberg exchange interaction is computed as follows:

$$\vec{H}_{exchange} = \frac{2A_{ex}}{\mu_0 M_s} \sum_i \frac{(\vec{m_k} - \vec{m_c})}{\Delta_i^2} \qquad (1)$$

Where $k$ represents the six nearest neighboring cells of the central cell, $\vec{m_c}$ is the magnetization of the central cell, $\vec{m_k}$ is the magnetization of $k^{th}$ neighboring cell, $\Delta_k$ is the cell size in the direction of neighbor $k$.

The effective field due to the perpendicular magnetic anisotropy, $\vec{H}_{anis}$ is given as:

$$\vec{H}_{anis} = \frac{2K_{u1}}{\mu_0 M_s} (\vec{z}.\vec{m})\vec{z}. \qquad (2)$$

Here, the first order uniaxial anisotropy constant is $K_{u1}$, the magnetic permeability of free space is $\mu_0$, and $\vec{z}$ is the unit vector corresponding to the anisotropy direction.

The temperature effect is calculated by,

$$\vec{H}_{thermal} = \vec{\eta}(step) \sqrt{\frac{2\alpha k_B T}{M_s \gamma \Delta V \Delta t}}$$

Where T is the temperature (K), $\Delta V$ is the cell volume, $k_B$ is the Boltzmann constant, $\Delta t$ is time step and $\vec{\eta}(step)$ is a random vector sampled from a standard normal distribution which is uncorrelated for each of the three cartesian coordinates and change value every time step.

The cell sizes chosen were 1 nm³, so that all dimensions are well within the limit of ferromagnetic exchange length calculated by, $\sqrt{2A_{ex}/\mu_0 M_s^2}$ = 4.86 nm, where the exchange stiffness, $A_{ex} = 15 \times 10^{-12}$ Jm$^{-1}$, saturation magnetization, $M_s = 1 \times 10^6$ Am$^{-1}$, and permeability of free space, $\mu_0 = 4\pi \times 10^{-7}$ H/m.

While PMA is created from the interaction between the ferromagnet's hybridized $d_{xz}$ and oxygen's $p_z$ orbital at a ferromagnet/oxide interface [1], by the application of voltage pulse, the interface electron density as well as perpendicular anisotropy can be changed [2]. This phenomenon is called VCMA [3-5].

The nanomagnets that induce magnetic field in the spin ensemble are elliptical in shape with easy axis, $D_{easy}$=36 nm, hard axis, $D_{hard}$=100 nm and the volume is calculated by, $\pi D_{easy} D_{hard} h = 900\pi$ nm³, which is the same volume as the skyrmion $\pi r^2 h = 900\pi$ nm³.

    The nanomagnets and the mesoscopic qubit are assumed to be in an external magnetic field along the direction of z-axis. The magnetization is initialized to out-of-plane direction due to PMA and global magnetic field ($B_0$). To alter the magnetizations of the two identical nanomagnets, VCMA is applied through the variation of PMA.

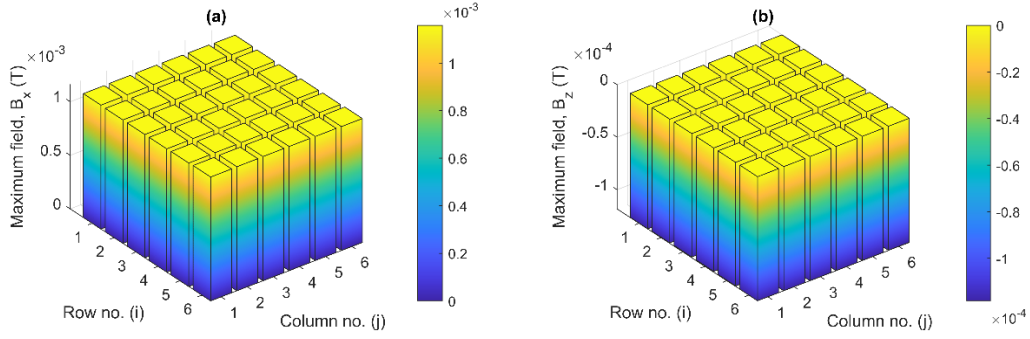

**Figure 1: Induced magnetic fiend ($B_x, B_z$) by nanomagnets.**

**Electron spin evolution:**

The Hamiltonian has a time-independent and a time-dependent part:

$$H(t) = H_0 + H_1(t)$$

$$H(t) = -(\omega_x^{st} + \omega_x^{var}(t))\widehat{S_x} - (\omega_y^{st} + \omega_y^{var}(t))\widehat{S_y} - (\omega_z^{st} + \omega_z^{var}(t))\widehat{S_z}$$

We assume a static field along the z-direction,

$$H(t) = -\omega_x^{var}(t)\widehat{S_x} - \omega_y^{var}(t)\widehat{S_y} - (\omega_0 + \omega_z^{var}(t))\widehat{S_z}$$

Here, $\omega_0 = \gamma B_0 \hat{z}$ is the angular velocity or Larmor frequency of electron spins when subjected to the external static field $B_0$.

The density matrix represents the linear form of Hilbert space of pure quantum states that represents a statistical ensemble of quantum states in a Bloch region. The density matrix can be written as the convex sum of pure state density matrices:

$$\rho(t) = \sum_i p_i |\psi_i\rangle\langle\psi_i|$$

Here, $p_i$ is the relative weight of the quantum state $|\psi\rangle$.

The Hamiltonian and density matrix operators can be written in the rotating frame of reference as:

$$H(t) = e^{i\omega_r S_z t} H(t) e^{-i\omega_r S_z t}$$

$$\rho(t) = e^{i\omega_r S_z t} \rho(t) e^{-i\omega_r S_z t}$$